\documentclass[aps,superscriptaddress, showpacs,preprintnumbers, superscriptaddress, nofootinbibt,onecolumn,14pt,reqno]{revtex4-2}
\usepackage{eurosym}
\usepackage{dcolumn}
\usepackage{bm}
\usepackage{enumerate}
\usepackage{float}
\usepackage{amsmath}
\usepackage{bm}
\usepackage{amsfonts}
\usepackage{amssymb}
\usepackage{graphicx}
\usepackage{alphalph,mathtools}
\usepackage{xcolor}
\def\be{\begin{equation}}
\def\ee{\end{equation}}
\def\bea{\begin{eqnarray}}
\def\eea{\end{eqnarray}}

\usepackage{multirow}
\begin{document}
\title{ Quintessences Universe  in $ f(R,L_{m}) $ gravity with special form of deceleration parameter}
\author{B. K. Shukla}
\email{bhupendrashukla695@gmail.com}
\affiliation{Department of Mathematics, Govt. College,  Bandri Sagar (M.P.) India}
\author{R. K. Tiwari}
\email{rishitiwari59@rediffmail.com}
\affiliation{Department of Mathematics, Govt. Model Science College, Rewa 486 001 (M.P.) India}
\author{D. Sofuo$ \widehat{g} $lu}
\email{degers@istanbul.edu.tr.}
\affiliation{Department of Physics, Istanbul University Vezneciler 34134, Fatih, Istanbul, Turkey}
\author{A. Beesham}
\email{abeesham@yahoo.com}
\affiliation{Department of Mathematical Sciences, University of Zululand,P Bag X1001,Kwa-Dlangezwa 3886, South Africa}
\affiliation{Faculty of Natural Sciences, Mangosuthu University of Technology, P O Box 12363, Jacobs, South Africa}
\affiliation{National Institute for Theoretical and Computational Sciences, South Africa}
\begin{abstract}
In this paper We have  investigated  a homogeneous and  isotropic FRW cosmological  model with perfect fluid in the framework  of $f (R, L_{m})$ gravity. We have explored for the non linear case   of $f (R, L_{m})$ model,  namely $f (R, L_{m}) = \frac{R}{2} + L_{m}$   and obtained the solution by using  the condition that   the deceleration parameter  is a  linear function of  the Hubble parameter. We employ 57 Hubble data points and 1048 Pantheon supernovae type Ia data samples to restrict the model parameters. Additionally, we employ Markoc Chain Monte Carlo (MCMC) simulation for our statistical analysis. Additionally, we analyse the jerk and om diagnostic parameters for our model using the parameter values that were obtained.
\end{abstract}

\pacs{2}
\date{\today }
\maketitle

\section{Introduction}
Numerous scientific investigations have been conducted in the last several decades to understand the mysterious behaviour of the universe. Throughout their entire existence, gravitational waves, dark energy, early time inflation, late time acceleration,BH ,WH and cosmological constants have been probing the very nature of the universe. The universe was thought to be both isotropic and homogenous on an enormous scale, just to investigate into the cosmological principle. But in 1992,  COBE convincingly asserted that the large-scale CMB contains a minor anisotropy \cite{1}. This was also confirmed by observations taken by the Plank collaborations \cite{2}, CBI \cite{3}, WMAP,\cite{4} and  BOOMERanG \cite{5} in following years. The observational findings of the two teams led by Perlmutter and Riess also led to exciting developments in the realm of cosmology \cite{6,7}. These investigations aim to support the idea that the cosmos is presently experiencing an accelerated expansion phase. The isotropic character of the universe's expansion has been a topic of debate up to this point. The universe appears to grow at a variable rate in other directions, according to interesting recent findings \cite{8}.Despite being the most successful, FLRW cosmology is founded on cosmic principles. \\
In the current instance, the modified theoretic technique seems more effective to deal with the examination of such problems. One of these is the $f (R)$ theory of gravity, which has created a solid framework to evaluate the present state of the universe's evolution \cite{9}. In fact, the interpretations of late-time acceleration\cite{10,11}, the exclusion of the dark matter entity in the examination of the dynamics of large test particles \cite{12}, and the unification of inflation with dark energy \cite{13} may all be satisfactorily explained by $f (R)$ theories. Numerous arguments also suggest that higher-order theories, such as $f (R)$ gravity, are able to account for the flatness of galaxies' rotating curves \cite{14}. Numerous coupling ideas were created as a result of these reasons\cite{15,16,17}.The $f (R, L_{m})$ theory of gravity is one of them \cite{18}. Notably, this favours the occurrence of an additional, four-velocity-orthogonal force. Additionally, the 'additional' force explains the test particle's non-geodesic movements.
As a result, it is possible to identify an equivalence principle failure. The literature has produced many improvements to this hypothesis \cite{19,20,21,22,23,24,25,26}. Jaybhaye et al. recently investigated cosmology under $f (R, L_{m})$ gravity \cite{27}.\\
In this paper, we have investigated homogeneous and spatially isotropic FRW model with a perfect fluid in this study. Within the constraints of $f (R, L_{m})$ gravity, this is succeeded in doing. We focus on a nonlinear $f (R, L_{m})$ model, namely $f (R, L_{m}) = \frac{R}{2} + L_{m}$. Furthermore, the deceleration parameter is a unique variation of the Hubble parameter. We employ 57 Hubble data points and 1048 Pantheon supernovae type Ia data samples   to restrict the model parameters. Additionally, we employ Markoc Chain Monte Carlo (MCMC) modelling for our statistical analysis. Additionally, we analyse the jerk and om diagnostic parameters for our model using the parameter values that were obtained.\\
The content of this paper is structured as follows: Section II presents the fundamental definition of $f (R, L_{m})$ gravity.
In section III, the study of deceleration parameter is a special form of Hubble parameter. The examination of observational conditions and a discussion of the findings are presented in Section IV.In section V, we explain the cosmographic parameter and om diagnostic in section VI. A few closing remarks are provided in section VII, the last section.

\section{Review of $ f(R,L_{m}) $ gravity}
The gravitational action for $f(R, L_{m})$ is given by
\bea
\mathcal{S}=\int f(R,L_{m})\sqrt{-g} dx^{4}.\label{eq1}
\eea
Here $ R $ denote the Ricci scalar and $ L_{m} $ denotes the matter Lagrangian .\\
Ricci scalar $ R $ is obtained by the following method 
\bea
R=g^{\mu\nu}R_{\mu\nu}, \label{eq2}
\eea
where the Ricci tensor $ R_{\mu\nu} $ is written in the following form as
\bea
R_{\mu\nu}=\partial_{\lambda}\Gamma^{\lambda}_{\mu\nu}-\partial_{\mu}\Gamma^{\lambda}_{\lambda\nu}+\Gamma^{\lambda}_{\mu\nu}\Gamma^{\sigma}_{\sigma\lambda}-\Gamma^{\lambda}_{\nu\sigma}\Gamma^{\sigma}_{\mu\lambda},\label{eq3}
\eea
where $ \Gamma^{\lambda}_{\alpha\beta} $ is the components of Levi-Civita
connection.\\
Now the given field equation is obtained by with respect to the metric tensor $ g_{\mu\nu} $  
\bea
f_{R}R_{\mu\nu}+(g_{\mu\nu}\square-\nabla_{\mu}\nabla_{\nu})f_{R}-\frac{1}{2}(f-f_{L_{m}}L_{m})g_{\mu\nu}=\frac{1}{2}f_{L_{m}}T_{\mu\nu}, \label{eq4}
\eea
Here $ f_{R}=\frac{\partial f}{\partial R} ,f_{L_{m}}=\frac{\partial f}{\partial L_{m}}$ and  $ T_{\mu\nu} $ is called  the stress-energy tensor for the cosmic fluid, given by
\bea
T_{\mu\nu}=\frac{-2}{\sqrt{-g}} \frac{\delta (\sqrt{-g}L_{m})}{\delta g^{\mu\nu}}.\label{eq5}
\eea
Now, we can obtain the following  relation by using covariant derivative in equation (\ref{4})
\bea
\nabla^{\mu}T_{\mu\nu}=2\nabla^{\mu} ln(f_{L_{m}})\frac{\partial L_{m}}{\partial g^{\mu\nu}}.\label{eq6}
\eea
To find the cosmological implications,  we consider the following homogeneous and spatially isotropic  FLRW metric 
\bea
ds^{2}=-dt^{2}+a^{2}(t)(dx^{2}+dy^{2}+dz^{2}),\label{eq7}
\eea
where, $a(t)$ is the cosmic scale factor. Now , the Ricci scalar  for the metric (\ref{eq7}) is obtained
\bea
R=6(\dot{H}+2H^{2}),\label{eq8}
\eea
where $ H $ is Hubble parameter and defined as
\bea
H=\frac{\dot{a}}{a}.\label{eq9}
\eea
The energy-momentum tensor for energy density $ \rho $ and the pressure $ p $ of the cosmic  pressure is taken by,
\bea
T_{\mu\nu}=(\rho+p)u_{\mu}u_{\nu}+pg_{\mu\nu}.\label{10}
\eea
The corresponding field equations  for above are given by
\bea
3H^{2}f_{R}+\frac{1}{2}(f-f_{R}R-f_{L_{m}}L_{m})+3H\dot{f}_{R}=\frac{1}{2}f_{L_{m}}\rho,\label{eq11}
\eea
\bea \label{eq12}
\dot{H}f_{R}+3H^{2}f_{R}-\ddot{f}_{R}-3H\dot{f}_{R}+\frac{1}{2}(f_{L_{m}L_{m}}-f)=\frac{1}{2}f_{L_{m}} p.
\eea
\section{Deceleration parameter in term of Hubble parameter}
In this section, we explore a variable deceleration parameter in order to find the exact solution of the field equations.To explain how the universe changed from an era of decelerating expansion to one of current acceleration, a variable deceleration parameter, q, has been suggested \cite{28,29,30,31,32,33,34}.\\
The cosmos expands either more accelerating $(q<0)$  or more decelerating $(q>0)$, depending on the geometric parameter known as the deceleration parameter.When $q=0$, the universe grows at a steady rate. When $q<-1$, the accelerated growth is referred to as super-exponential expansion. A new variable deceleration parameter has been assumed with the Hubble parameter as \cite{35}
\bea
q=\alpha -\frac{\beta}{H^{2}},\label{eq17}
\eea
where the positive constants $ \alpha $ and $ \beta $ are present. Using the formula for the Hubble parameter $ H=\frac{\dot{a}}{a} $,and the deceleration parameter in equation (13) is $ q=-1-\frac{\dot{H}}{H^{2}} $. We can determine the scale factor $a$
\bea
a=[\sinh\tau]^{\frac{1}{1+\alpha}}.\label{eq18}
\eea
where $ \tau =\lbrace(\sqrt{(1+\alpha)\beta)}t+c\rbrace $ and  $c$ is an integration constant. The Hubble parameter $H$ and the deceleration parameter $q$ are rewritten by the value of (\ref{eq18})
\bea
H=\sqrt{\frac{\beta}{1+\alpha}}\coth\tau,\label{eq19}
\eea
\bea
q=\alpha-(1+\alpha)\tanh^{2}\tau,\label{eq19}.
\eea
Furthermore, we may express the Hubble parameter and deceleration parameter in terms of redshift by using the relationship between the redshift and the scale factor of the universe, $a(t) =(1+z)^{-1}$ .
\bea
H=\sqrt{\frac{\beta}{1+\alpha}}[1+(1+z)^{2(1+\alpha)}]^{\frac{1}{2}},\label{eq20}
\eea
\bea
q=\alpha -\frac{(1+\alpha)}{[1+(1+z)^{2(1+\alpha)}]}.\label{eq21}
\eea
\section{Evaluation to model parameters according to observational data}
The Hubble parameter for redshift, which has been given by
\bea
H=\sqrt{\frac{\beta}{1+\alpha}}[1+(1+z)^{2(1+\alpha)}]^{\frac{1}{2}},\label{eq22}
\eea
Using the Hubble parameter's most recent value, determine the parameter
\bea
H=\frac{H_{0}}{\sqrt{2}}[1+(1+z)^{2(1+\alpha)}]^{\frac{1}{2}},\label{eq23}
\eea
where $ H_{0}=\sqrt{\frac{2\beta}{(1+\alpha)}} $.
\subsection{Cosmic Chronometers datasets}
We then employ 57 Hubble data sets in the redshift region of $0.07 < z <2.36$ offered by Magana et al. \cite{36} to determine the model parameter and the latest value of the Hubble parameter $H_{0}$.\\
The associated chi-square function is written as such to determine the mean value of the parameters and $H_{0}$:
\bea
\chi^{2}_{H}(\alpha,H_{0})=\displaystyle \sum_{i=1}^{57}\frac{[H_{th}(\alpha,H_{0},z_{i})-H_{obs}(z_{i})]^{2}}{\sigma^{2}_{i}},\label{eq24}
\eea
where $H_{th}(\alpha,H_{0},z_{i})$ and $ H_{obs}(z_{i}) $, respectively, represent the model-predicted and observed Hubble parameter measurements. The standard error of the Hubble parameter measurement is denoted by the symbol $ \sigma_{i} $.
\subsection{Pantheon datasets}
In this some cases, we use 1048 points from the Pantheon Type 1a supernova data set (Scolnic et al.\cite{37}) to estimate the parameters $ \alpha $ and $ H_{0} $. This data collection contains 1048 apparent magnitude observations $ m_{B} $ that support the redshift range of 0$ 0.01 <z <2.3$.\\
In conventional cosmology, the distance modulus $\mu(z)$ is defined as
\bea
\mu=m_{B}-M=5 log_{10}(d_{l}(z))+25,\label{eq25}
\eea
where $ m_{B} $ is the perceived magnitude, $M$ is the absolute magnitude, and $ d_{l}(z) $ is the luminosity distance for the flat FRLW universe
\bea
d_{l}(z)=\frac{c(1+z)}{H_{0}}\int_{0}^{z}\frac{dz^{*}}{E(z^{*})}.\label{eq26}
\eea
Given that all supernovae are thought to have the identical actual size, we may use supernova 2002 cr, whose  possesses a $ m_{B} $ value of $13.907\pm 0.1982$ at a low redshift of $z = 0.0101$, to calculate the actual size $ M$  using the method described by Goswami et al. \cite{38}.
\bea
M= 5 log_{10}\left( \frac{H_{0}}{c}\right)-1.093.\label{eq27} 
\eea
Using equations (\ref{eq25}) and (\ref{eq27}), we arrive at the apparent magnitude equation shown below.
\bea
m_{B}= 5 log_{10}\left( \frac{d_{l}(z)H_{0}}{c}\right)+23.907.\label{eq28} 
\eea
Now,we define  chi-square function is 
\bea
\chi^{2}_{Pan}(\alpha,H_{0})=\displaystyle \sum_{i=1}^{1048}\frac{[\mu_{th}(\alpha,H_{0},z_{i})-\mu_{obs}(z_{i})]^{2}}{\sigma^{2}_{i}},\label{eq29}
\eea
where $ \mu_{th}(\alpha,H_{0},z_{i}), \mu_{obs}(z_{i}) $ and $ \sigma_{i}^{2} $ are theoretical distance , observational distance modulus and variance at $ z_{i} $ respectively.
\subsection{An observational finding}
We are capable to obtain limits on the parameters of my cosmological model for all of the two distinct situations for the combined $ H(z)$+Pantheon data set by optimising the complete chi-squared function$ \chi_{H}^{2}+\chi_{Pan}^{2} $.
\begin{center}
\includegraphics[width = 95 mm]{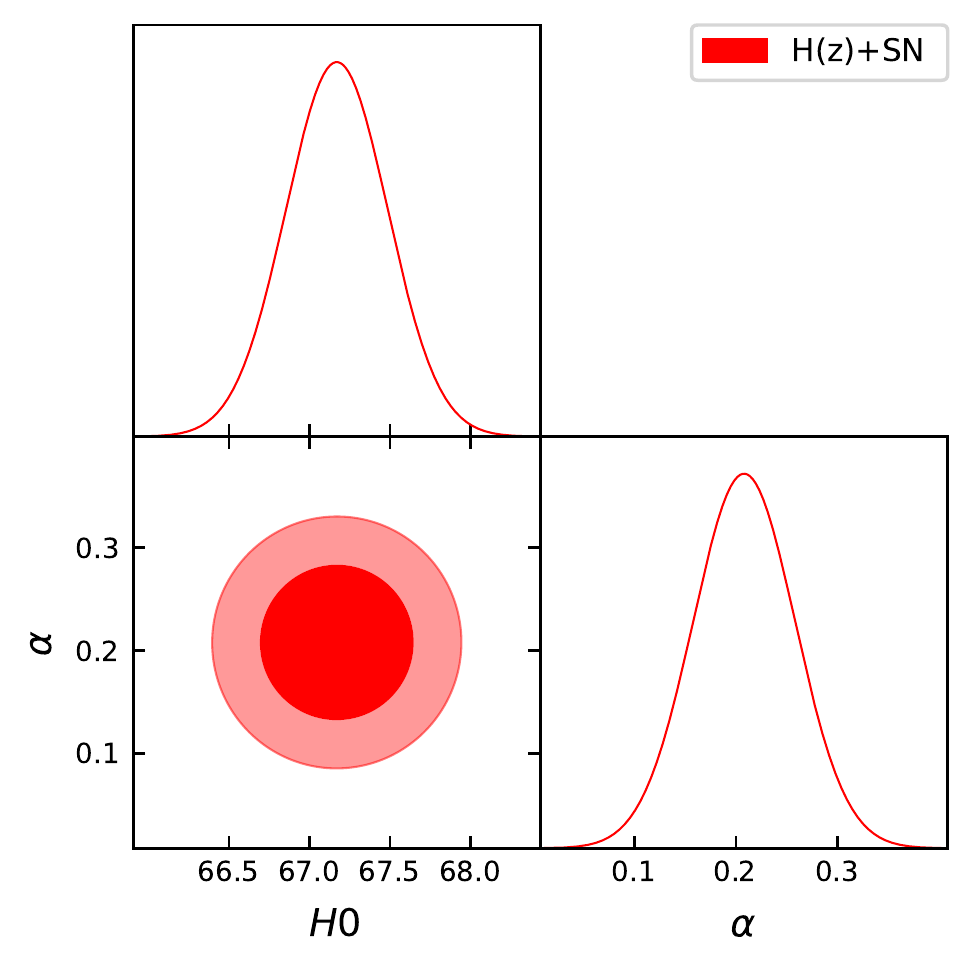}\\
\textbf{Fig-1.} The MCMC likelihood curves at $ 1\sigma $ and $ 2\sigma $ obtained from the CC+SN datasets are shown in the above figure.
\end{center}
\begin{center}
\includegraphics[width = 80 mm]{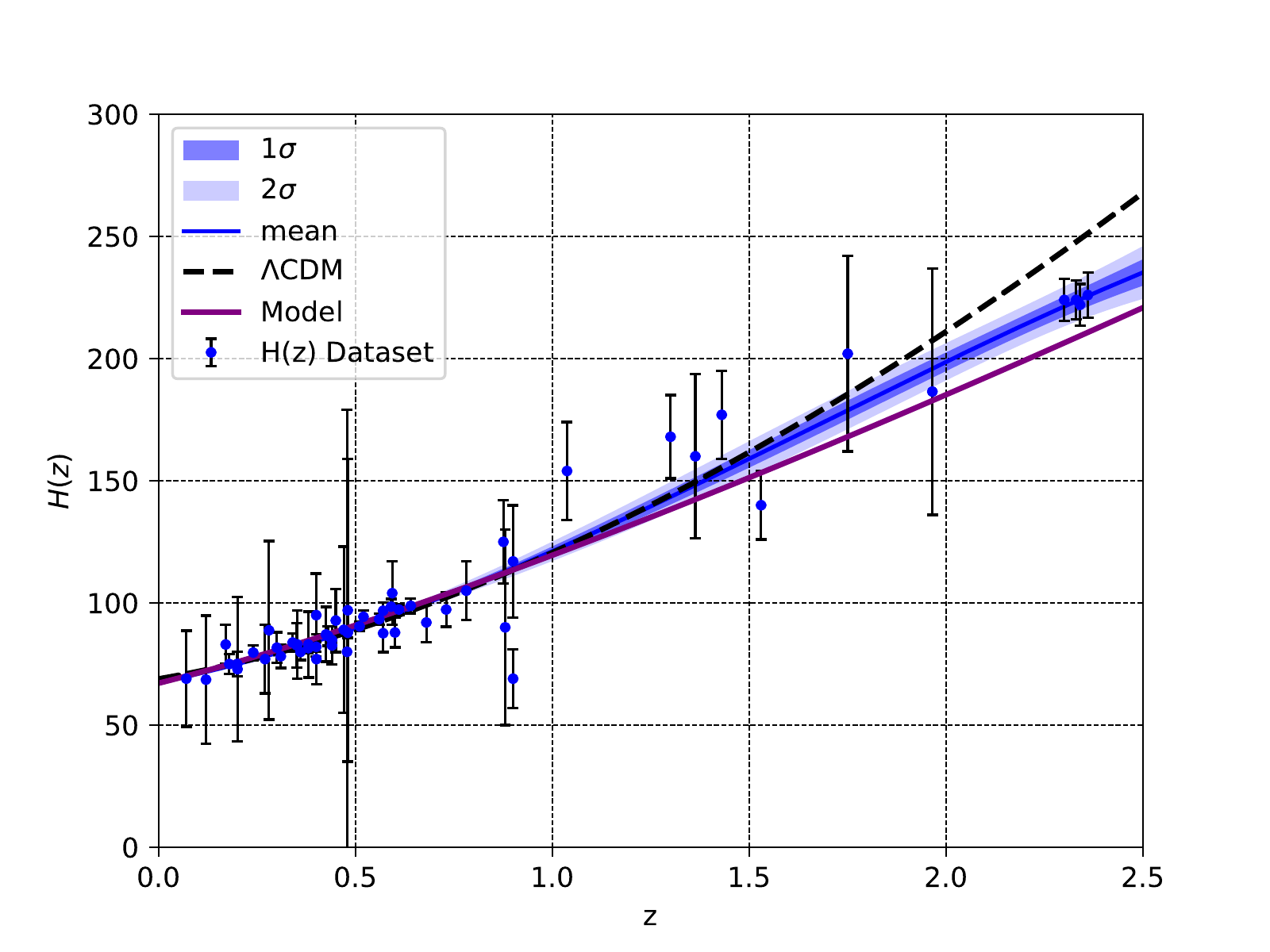}\hspace{1cm}\includegraphics[width = 80 mm]{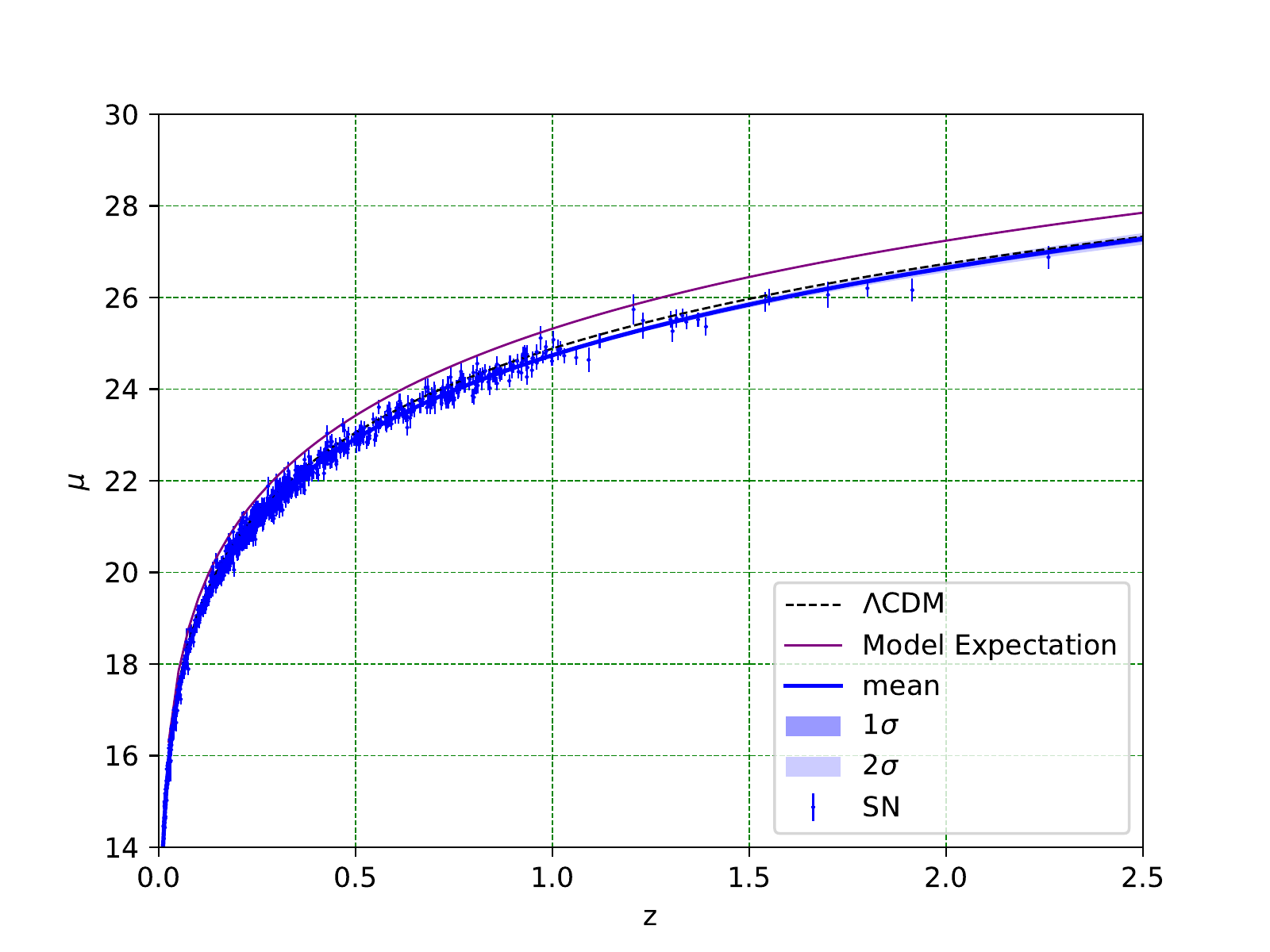}\\
\small \textbf{Fig-2.} The figure show that the theoretical curve of $ H(Z) $.\hspace{0.5cm}\textbf{Fig-3.} The figure show that the theoretical curve of $ \mu(z) $.
\end{center}
\begin{center}
Table 1 is an overview of the MCMC findings from the paper.
\end{center}
\begin{center}
\begin{tabular}{|c|c|c|}
\hline
\textbf{Observations data name} & \textbf{Parameters}& $ F(R,L_{m}) $ model\\
\hline
CC& $ \alpha $, $ H_{0} $ &  $ 0.208573\pm 0.011896 $, $67.166192\pm 0.401849$\\
\hline
SN& $ \alpha $, $ H_{0} $ &  $ 0.208573\pm 0.026998 $, $67.166192\pm 0.7202$\\
\hline
\end{tabular}
\end{center}
\section{ Cosmographic analysis}
In this section of our examination of the evolution of the universe, we employ a method known as "cosmography". It is predicated on the cosmological principle, which holds that our physical universe is homogeneous and isotropic on infinite sizes. Through this research, we are able to learn more about the evolution of the universe independent of the cosmological principle without supposing anything about certain cosmological theories. It makes advantage of the Taylor series' scale factor expansion
\bea
\frac{a(t)}{a(t_{0})}=1+H_{0}(t-t_{0})-q_{0}H_{0}^{2}\frac{(t-t_{0})^{2}}{2!}+j_{0}H_{0}^{3}\frac{(t-t_{0})^{3}}{3!}-......,\label{eq34}
\eea
Here, subscript $ 0 $ denotes the present value of the parameter.\\
We define Hubble parameter, deceleration parameter and jerk parameter in term of redshift as 
\bea
H(z)=-(1+z)^{-1}\frac{dz}{dt},\label{eq35}
\eea
\bea
q(z)=-1+\frac{(1+z)\frac{dH}{dz}}{H^{2}}.\label{eq36}
\eea
\bea
j(z)=(1+z)\frac{dq}{dz}+q(1+2q).\label{eq37}
\eea
We obtained the value of jerk parameter in terms of redshift
\bea
j(z)=\frac{2(1+\alpha)^{2}(1+z)^{2(1+\alpha)}}{[1+(1+z)^{2(1+\alpha)}]^{2}}+\left[\alpha -\frac{(1+\alpha)}{[1+(1+z)^{2(1+\alpha)}]} \right]+2\left[ \alpha -\frac{(1+\alpha)}{[1+(1+z)^{2(1+\alpha)}]}\right]^{2}  .\label{eq38}
\eea
\begin{center}
\includegraphics[width = 80 mm]{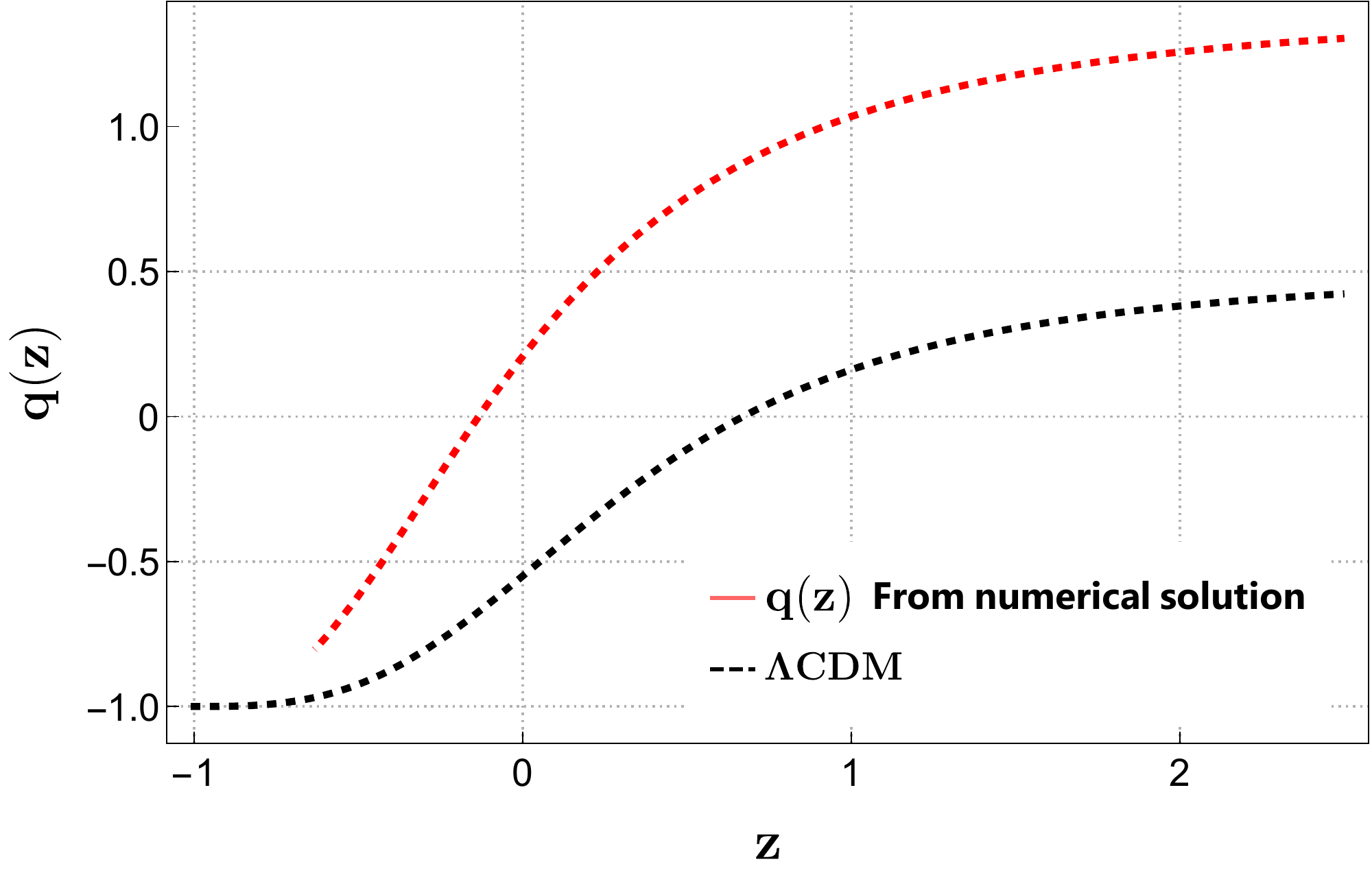}\hspace{0.5cm}\includegraphics[width = 80 mm]{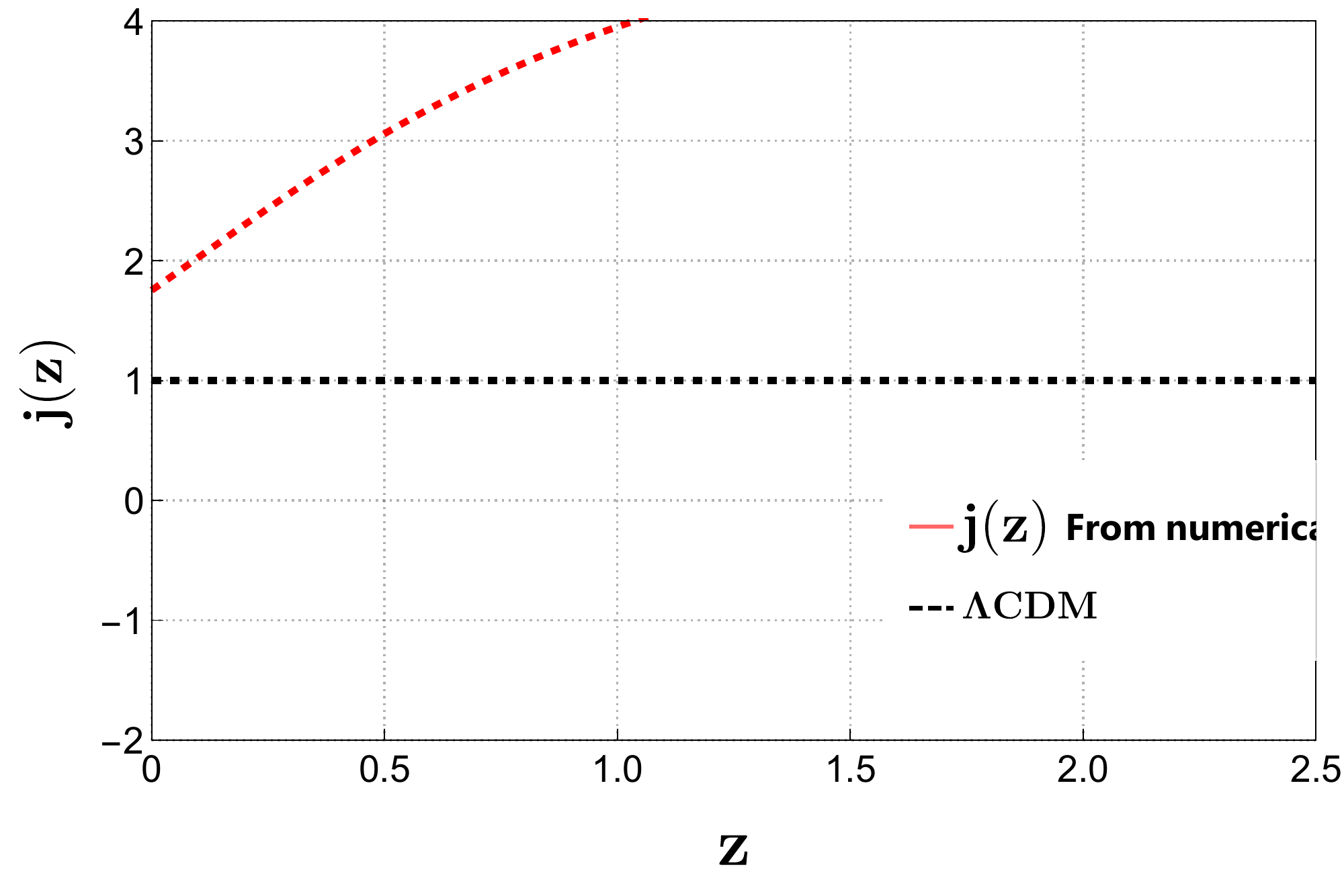}\\
\textbf{Fig-4.} The above figure shows the $ q $ vs $ z $.\hspace{2cm}\textbf{Fig-5.} The above figure shows the $ j$ vs $ z $.
\end{center}
\begin{center}
\includegraphics[width = 70 mm]{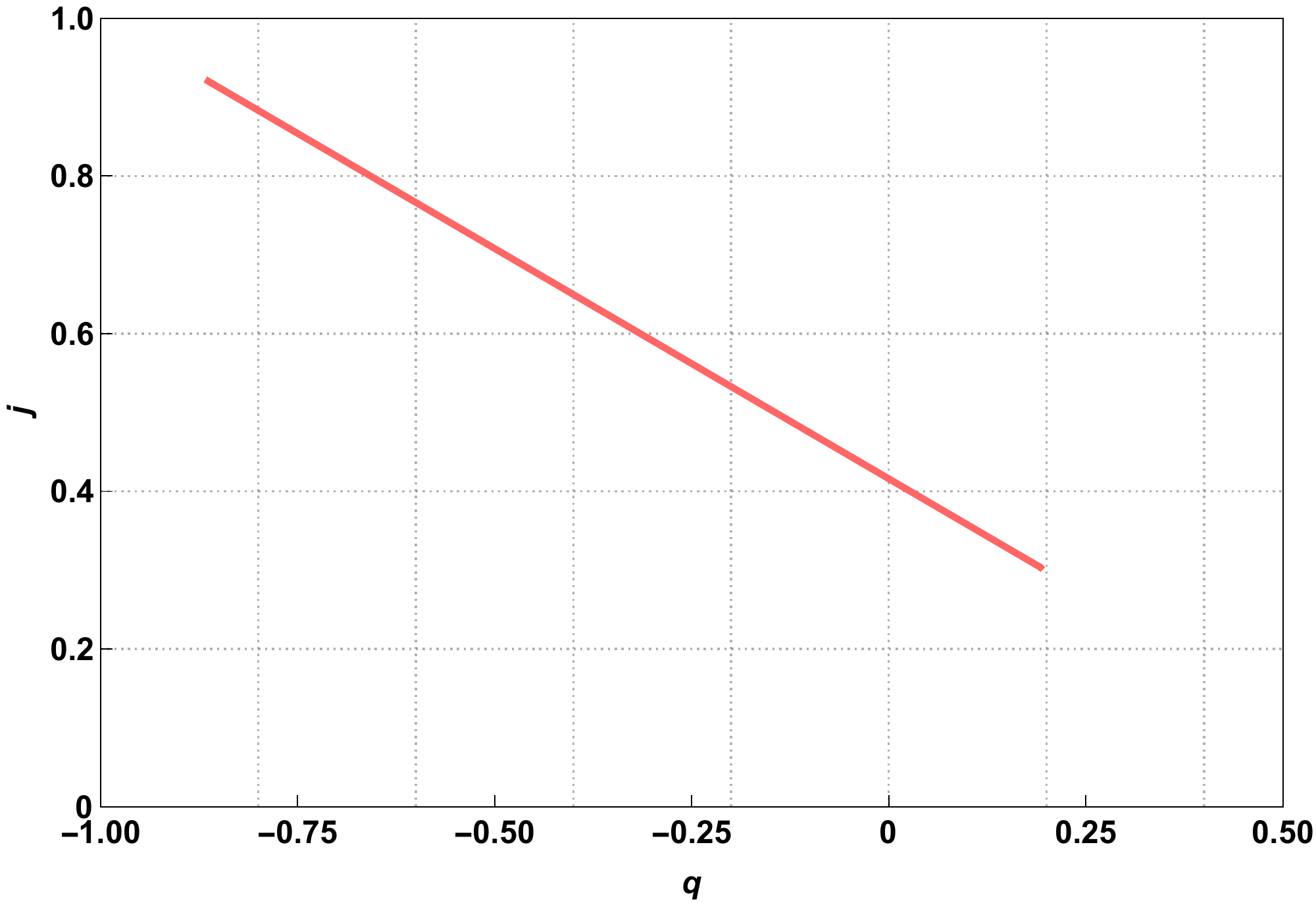}\\
\small \textbf{Fig-6.} The figure show that the deceleration parameter vs jerk parameter.
\end{center}
This adequately accounts for the observations of the standard $ \Lambda $CDM model. Figures 4, 5, and 6, which express the evolution of the jerk parameter and the deceleration parameter, respectively, also explain the characteristics of these parameters.
\subsection{Statefinder parameters}
Current observational data cannot rule out all of the cosmological dark energy concepts that are currently under consideration. To differentiate between alternative dark energy models, however, two new cosmological parameters were devised a few years ago. The scale factor and its derivatives with respect to cosmic time are used to express these so-called statefinder parameters completely up to the third order\cite{38}. The statefinder's parameters are described below
\bea
r=\frac{\dddot{a}}{aH^{3}},
\eea
\bea
s=\frac{r-1}{3(q-\frac{1}{2})}.
\eea
The characteristics of the plane's paths can be extremely different for several existing models. The departure of these trajectories from the (1,0) point determines the separation between a particular model and the LCDM model. The statefinder pair $(r-s)$ successfully separates interacting dark energy theories, brane models, quintessence, and Chaplygin gas from a variety of other cosmological models. For a particular model, it is possible to determine the pair $(r, s)$ and the trajectory in the $(r, s)$ plane.
\begin{center}
\includegraphics[width = 70 mm]{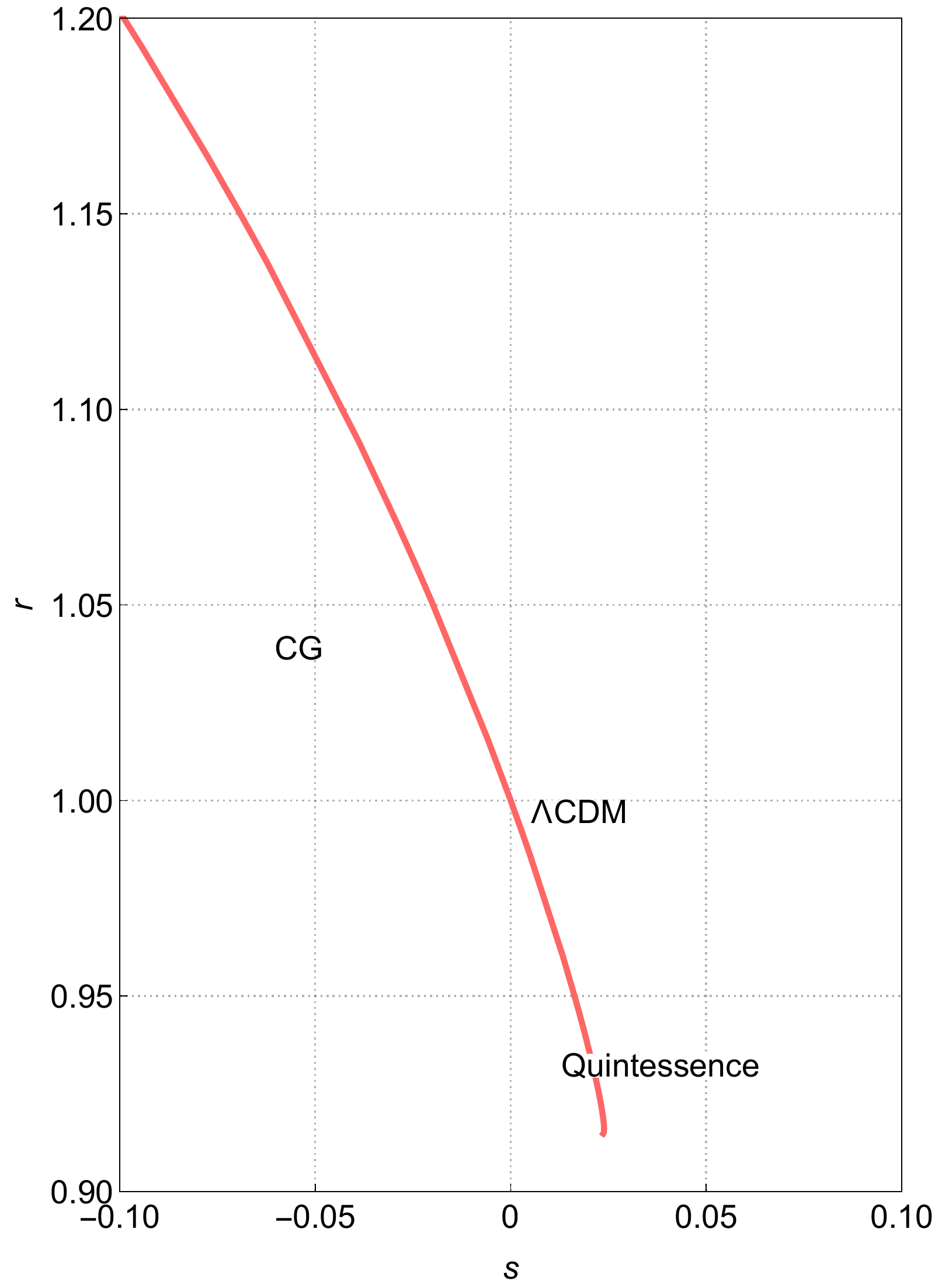}\\
\small \textbf{Fig-7.} The figure show that the statefinder parameter.
\end{center}
\section{Om DIAGNOSTIC $ Om(z) $}
Om is a mathematical diagnostic which includes the redshift and the Hubble parameter. With and without consideration of matter density, it can distinguish between a dynamical dark energy model and the CDM. Dark energy is a cosmological constant ($ \Lambda $CDM) because $Om(z)$ behaves consistently with regard to $z$. $Om(z)$ has a positive slope, which suggests that dark energy is phantom, and a negative slope, which suggests that dark energy behaves like quintessence. The definition of $Om(z)$ for a spatially flat Universe is given by
\bea
Om(z)=\dfrac{\left[ \frac{H(z)}{H_{0}}\right]^{2}-1 }{(1+z)^{3}-1},
\eea
where $ H_{0} $ is initial value of Hubble parameter.
\begin{center}
\includegraphics[width = 70 mm]{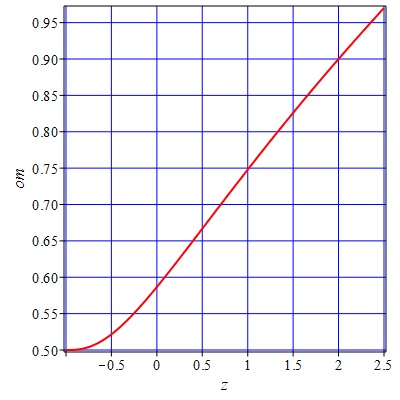}\\
\small \textbf{Fig-8.} The figure show that the om diagnostics.
\end{center}
\section{ $f(R, L_{m})$ gravity MODEL}
To examine the behaviour of the universe's, we determine this $f(R, L_{m})$ gravity.
\bea
f(R,L_{m})=\frac{R}{2}+L_{m}^{n},\label{eq36}
\eea
where $ n $ is free parameter.\\
Following that, for that particular functional type of $L_{m} = \rho $ \cite{40}, the universe is given by the Friedmann equations (\ref{eq11}) and (\ref{eq12})
\bea
3H^{2}=(2n -1)\rho^{n},\label{eq37}
\eea
\bea
2\dot{H}+3H^{2}=[(n-1)\rho-n p)\rho^{n-1}.\label{eq38}
\eea
the energy density and cosmic pressure in term of red shift
\bea
\rho =\left[ \frac{H_{0}^{2}[1+(1+z)^{2(1+\alpha)}]}{2(2n-2)} \right]^{\frac{1}{n}},
\eea
\bea
p=-\frac{1}{n} \dfrac{[-H_{0}^{2}(1+z)^{2(1+\alpha)}]+\frac{3}{2}H_{0}^{2}[1+(1+z)^{2(1+\alpha)}]}{\left[ \frac{H_{0}^{2}[1+(1+z)^{2(1+\alpha)}]}{2(2n-1)} \right]^{\frac{n-1}{n}}}+\frac{(n-1)}{n}\left[ \frac{H_{0}^{2}[1+(1+z)^{2(1+\alpha)}]}{2(2n-1)} \right]^{\frac{1}{n}},
\eea
\bea
\omega=\dfrac{-\frac{1}{n} \dfrac{[-H_{0}^{2}(1+z)^{2(1+\alpha)}]+\frac{3}{2}H_{0}^{2}[1+(1+z)^{2(1+\alpha)}]}{\left[ \frac{H_{0}^{2}[1+(1+z)^{2(1+\alpha)}]}{2(2n-1)} \right]^{\frac{n-1}{n}}}+\frac{(n-1)}{n}\left[ \frac{H_{0}^{2}[1+(1+z)^{2(1+\alpha)}]}{2(2n-1)} \right]^{\frac{1}{n}}}{\left[ \frac{H_{0}^{2}[1+(1+z)^{2(1+\alpha)}]}{2(2n-1)} \right]^{\frac{1}{n}}}.
\eea
\begin{center}
\includegraphics[width = 80 mm]{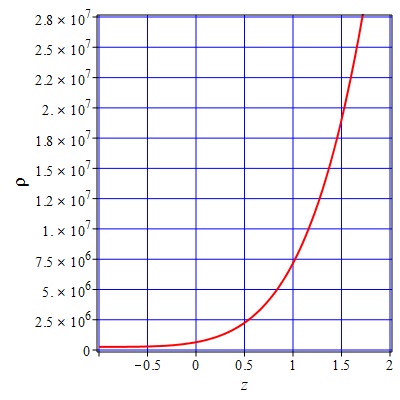}\hspace{0.5cm}\includegraphics[width = 80 mm]{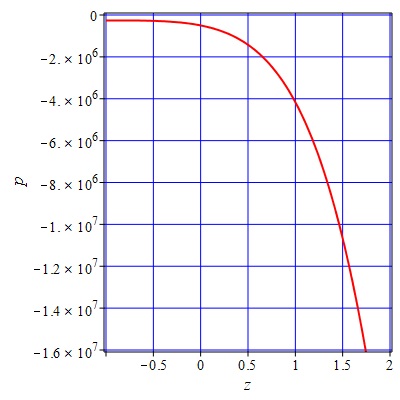}\\
\textbf{Fig-9.} Energy density $ \rho $ vs redshift $ z $.\hspace{2cm}\textbf{Fig-10.} Cosmic pressure $ p$ vs redshift $ z $.
\end{center}
\begin{center}
\includegraphics[width = 80 mm]{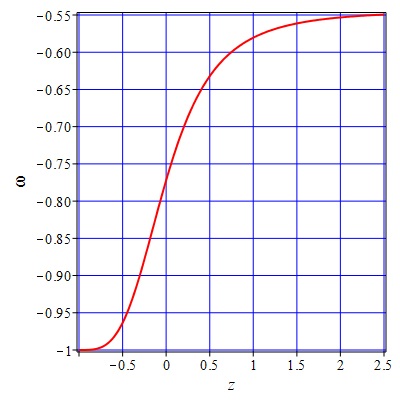}\\
\textbf{Fig-11.}The EoS  $ \omega $ vs redshift $ z $.
\end{center}
This figure 9 demonstrates that the energy density is a positive function of z and increases as cosmological redshift increases. When $z \rightarrow -1$, it begins with a positive value and moves towards zero. In Fig.10, which depicts the pressure behaviour as a function of redshift, we can see that the pressure in the current model decreases as the cosmological redshift increases. It initially has a high negative value and is currently approaching zero. Recent observations indicate that the so-called dark energy, which has a negative pressure, is the cause of the Universe's accelerated expansion phase. As a result, the pressure for our model fits recent data.\\
The EoS parameter is a crucial instrument for describing the cosmic epochs and comprehending the properties of dark energy. Each different dark energy model has a different range of values for this parameter. If the cosmological constant ($ \Lambda $CDM) characterises dark energy, then $\omega =- 1$. While if $-1<\omega<- 0.33$, we claim that dark energy is quintessential, and if $\omega<- 1$, the model's phantom character is indicated. Our model equation of state parameter lies between -1 and -0.55.
\section{Conclusion}
By adopting the novel relation of deceleration parameter  and   Hubble parameter in the non-linear f(R,Lm) formalism for FRW cosmological model. We have analyzed  the  Hubble parameter $H_{0}$ by  using 57 Hubble data points in the redshift range $0.07 < z < 2.36$   and  we use 1048 points from the Pantheon Type 1a supernova data to estimate the parameters $ \alpha $ and $H_{0}$. This data collection contains 1048 apparent magnitude observations $m_{B}$ that support the redshift range of $00.01 < z < 2.3$.The matter density is a decreasing function of time. The behavior of model is  closed to Lambda CDM model . The statefinder parameters  $(r s)$ indicate for  dark energy theories, brane models, quintessence, and Chaplygin gas .The EoS parameter is a crucial instrument for describing the cosmic epochs and comprehending the properties of dark energy. Each different dark energy model has a different range of values for this parameter. If the cosmological constant ($ \Lambda $CDM) characterises dark energy, then $\omega =- 1$. While if $-1 < \omega < -0.33$, we claim that dark energy is quintessential, and if $\omega < -1$, the model's phantom character is indicated. Our model equation of state parameter lies between -1 and -0.55.\\
\textbf{Abbreviations}\\
The following abbreviations are used in this manuscript:\\
BH:black holes,\\
WH: wormholes\\
COBE:Cosmic Background Explorer \\
CMB:cosmic microwave background\\
CBI:Cosmic Background Imager\\
WMAP:Wilkinson Microwave Anisotropy Probe\\
BOOMERanG:Balloon Observations of Millimetric Extragalactic Radiation and Geophysics\\


\begin{thebibliography}{1}
\bibitem{1} C. L. Bennett, A. J. Banday, K. M. Gorski, ´ ApJ, 464, L1,
(1996).
\bibitem{2} N. Aghanim, Y. Akrami, M. Ashdown et al., A $\&$ A, 641,
A6, (2020).
\bibitem{3} P. de Bernardis, P. A. R. Ade, J. J. Bock et al., BOOMERanG
Collaboration, AIP Conf. Proc., 555, 85, (2001).
\bibitem{4} B. S. Mason et al, ApJ, 591, 540, (2003).
\bibitem{5} G. Hinshaw, D.Larson, E. Komatsu et al., ApJS, 208, 19,
(2003).

\bibitem{6} S. Perlmutter et al, Bull.Am.Astron.Soc., 29, 1351, (1997).
\bibitem{7} A. G. Riess, A. V. Filippenko1, P. Challis et al., Astron J,
116, 1009, (1998).
\bibitem{8} K. Migkas , G. Schellenberger , T. H. Reiprich et al, A $\&$ A,
636,A15, (2020),.

\bibitem{9} A. Starobinsky, Phys. Lett. B, 91, 99, (1980).
\bibitem{10} S.M. Carroll, V. Duvvuri, M. Trodden, M.S. Turner, Phys.
Rev. D, 70, 043528, (2004).
\bibitem{11} S. Capozziello, V.F. Cardone, A. Troisi, Mon. Not. R. Astron. Soc., 375, 1423, (2007).
\bibitem{12} S. Nojiri, S.D. Odintsov, Phys. Lett.B, 657, 238, (2007).
\bibitem{13} S. Capozziello, V. F. Cardone, S. Carloni, A. Troisi, Phys.
Lett. A, 326, 292, (2004).
\bibitem{14} T. Harko, F. S. N. Lobo, S. Nojiri, S. D. Odintsov, Phys. Rev.
D, 84, 024020, (2011).
\bibitem{15} M. J. S. Houndjo, et al., Int. J. Mod. Phys. D, 26, 1750024,
(2017).
\bibitem{16} S. D. Odintsov, D. Saez-G’omez, ´ Phys. Lett. B, 725, 437,
(2013).
\bibitem{17} T. Harko, F. S. N. Lobo, Eur. Phys. J. C, 70, 373, (2010).
\bibitem{18} O. Bertolami., J. Paramos, S. Turyshev, arXiv:grqc/ 0602016, (2006).
\bibitem{19} O. Bertolami, C. G. Bohmer, T. Harko, F. S. N. Lobo, Phys.
Rev. D, 75, 104016, (2007).
\bibitem{20} T. Harko, Phys. Lett. B, 669, 376, (2008).
\bibitem{21} T. Harko, Phys. Rev. D, 81, 044021, (2010).
\bibitem{22}] J. Wang, K. Liao, Class. Quantum Grav., 29, 215016, (2012).
\bibitem{23} B.S. GonA§alves, P.H.R.S. Moraes, ˜ arXiv:2101.05918v1,
(2021).
\bibitem{24} G.A. Carvalho etal., Eur. Phys. J. C, 80, 483, (2020).
\bibitem{25} L. V. Jaybhaye, S. Mandal, P. K. Sahoo, IJGMMP, 19, 04,
(2022).
\bibitem{26} F. S. N. Lobo and T. Harko, arXiv:2203.03295v1, (2022).
\bibitem{27} L. V. Jaybhaye, R. Solanki, S. Mandal, P.K.Sahoo, Phys.
Lett. B, 831, 137148, (2022)
\bibitem{28} R.K. Tiwari, A. Beesham,B. Shukla,;   Int. J.
Geom. Meth. Mod. Phys.15, 1850115 (2018).
\bibitem{29} R.K. Tiwari, A. Beesham,B.K. Shukla,; Eur. Phys. J.
Plus 2016, 131, 447–456. [CrossRef]
\bibitem{30}  R.K. Tiwari, A. Beesham,B.K. Shukla,;  Eur. Phys. J.
Plus ,132, 20 (2017) . 
\bibitem{31}  R.K. Tiwari, A. Beesham,B.K. Shukla,;  Eur. Phys. J. Plus,132, 126 (2017).
\bibitem{32}  R.K. Tiwari, A. Beesham,B.K. Shukla,;  Int. J. Geom.
Methods Mod. Phys. 15, 1850155 (2018).
\bibitem{33} R.K.Tiwari, D. Sofuoglu, R. Isik,B.K. Shukla, E. Baysazan,; Int. J.Geom. Meth. Mod. Phys. 19, 2250118 (2022) . 
\bibitem{34} R.K.Tiwari, D. Sofuoglu,V.K. Dubey,;  Int. J. Geom.
Meth. Mod. Phys.17, 2050187 (2020).
\bibitem{35} R.K. Tiwari, et al;Symmetry,15, 788 (2023) .
\bibitem{36} J. Magana, M.H. Amante,M.A. Garcia-Aspeitia, V. Motta,  Mon. Not. R. Astron. Soc. 476(1), 1036(2018). 
\bibitem{37} D.M. Scolnic,  et al.,  Astrophys. J. 859(2), 101 (2018).
\bibitem{38} G.K. Goswami,   Res. Astron. Astrophys. 17(3),27(2017).
\bibitem{39} V. Sahni, T.D. Saini, A.A. Starobinsky, U. Alam
JETP Lett., 77,201 (2003). 
\bibitem{40}  T. Harko,;  F.S.N.Lobo,;J.P. Mimoso, ; D. Pavon,;  Eur. Phys. J. C, 75, 386( 2015)
\end{thebibliography}
\end{document}